\documentclass[11pt,twoside]{article}
\usepackage{FUSE2004}

\usepackage{epsf}
\usepackage{psfig}

\begin{document}
\title{Can Dust Explain Variations in the D/H Ratio?}
\author{B. T. Draine}
 
\affil{Institute for Advanced Study, Einstein Dr., Princeton, NJ 08540 USA;
 permanent address: Princeton University, Princeton, NJ 08544, USA}

\begin{abstract}
The D/H ratio in interstellar gas varies on scales of a few
hundred pc in the local Milky Way, with D/H values ranging from $\sim$7$\,$ppm
to $\sim$22$\,$ppm.
The reduction in D/H relative to the
primordial value of D/H$\,\approx26\,$ppm 
is usually
attributed to ``astration'' -- conversion of D into other elements by
nuclear fusion in stars.
However, it is shown here that
astration has difficulty accounting for the observations
because the expected associated variations in O/H are not seen.

The lower D/H values are instead likely 
due to ``depletion'' of the D onto dust grains.
Polycyclic aromatic hydrocarbons (PAHs)
are a possible repository for the missing D,
and it appears possible for gas-grain reactions to achieve extreme
deuteration of carbonaceous grain material.
Grain destruction will release D from the grains; the
gas phase abundance of D should therefore be positively correlated with
the gas phase abundances of other elements that
exhibit strong depletions, such as Mg, Si, Ti, and Fe,
which also will be returned to the gas by
grain destruction.
\end{abstract}

\section{Introduction}

In recent years it has become clear that the gas-phase D/H ratio
varies significantly from one sightline to another within a few hundred
pc of the Sun
(see, e.g., Moos et al.\ 2002; Steigman 2003; Wood et al.\ 2004;
H\'ebrard 2005).  
This has often been taken to result from regional
variations in the extent to which nucleosynthesis in stars, followed by
return of matter to the interstellar medium via stellar winds and ejecta,
has lowered the interstellar D/H ratio from
the primordial value established by nucleosynthesis in the Big Bang.

I argue here that the observed local variations in the gas-phase D/H ratio are
not due to variable astration.
The low values of (D/H)$_{\rm gas}$ seen on some sightlines are instead
attributed
to incorporation of D into dust grains.  
I discuss 
processes acting to add D to dust grains, and show that
these can strongly deuterate PAH material in the interstellar medium.
This hypothesis predicts 
a positive correlation of (D/H)$_{\rm gas}$
with gas-phase
abundances of strongly-depleted metals, and
deuteration of 
carbonaceous interstellar grains captured by the {\it Stardust} mission.

\section{D/H Variations in the ISM}

Using high resolution spectroscopy with IMAPS, Jenkins et al.\ (1999)
measured 
D/H=$7.4_{-0.9}^{+1.2}\,$ppm toward $\delta$~Ori ($d=281\pm65\,$pc), and
Sonneborn et al.\ (2000) found
D/H=$21.8_{-1.9}^{+2.2}\,$ppm toward
$\gamma^2$~Vel ($d=258\pm35\,$pc).
These two high precision measurements
firmly established the reality of substantial variations in the
gas-phase D/H ratio within a few hundred pc of the Sun.
Including observations with Copernicus, IMAPS, HST, and FUSE
(see the summary by Wood et al.\ 2004) we now
have measurements of D/H on $\sim$40 interstellar sightlines, with
D/H values ranging from $5.0\pm1.6\,$ppm toward $\theta$~Car 
(Allen, Jenkins, \& Snow 1992)
to $21.8_{-1.9}^{+2.2}\,$ppm toward $\gamma^2$~Vel 
-- a factor of $\sim$4 variation in D/H.

\subsection{Problems with Astration}

The gas returned to the ISM from stars is expected to be nearly devoid of D,
as the D is converted to $^3$He 
during the pre-main-sequence
evolution of $M\la 5M_{\sun}$ stars (Mazzitelli \& Moretti 1980).
As a result, the D/H ratio in the ISM is an
indicator of the fraction of the baryons now in the ISM 
that have passed through a star.
If the primordial value of D/H is $26.2_{-2.0}^{+1.8}\,$ppm 
(Spergel et al.\ 2003) then a
D/H ratio of $5.0\pm1.6\,$ppm 
would require that $81\pm6\%$ of
the baryons in the interstellar gas toward $\theta$~Car have been
cycled through a star at least once, whereas on the sightline toward
$\gamma^2$~Vel, this fraction would be only $17_{-11}^{+9}\%$.

The hypothesis that the observed variations in D/H on different sightlines
are due to variations in astration therefore requires that (1) regions
separated by only a few hundred pc have had extremely different star formation
histories, and (2) turbulent diffusion has not homogenized the
elemental abundances in the gas.  It is difficult to see how these
two conditions can be true: we do not see other evidence of such
variations in star formation activity, and turbulent mixing is expected to
be effective at mixing gas over length scales of hundreds of pc on 
$10^9\,$yr time scales.

The gas returned to the ISM from stars will, on average, be
enriched in the products of stellar nucleosynthesis, such as oxygen.
Models for evolution of interstellar abundances due to stellar nucleosynthesis
and infall predict the joint variation of D/H and O/H.
Outside of dark clouds, H$_2$O is absent, and the O in grains
is thought to be mostly in the form of amorphous olivine 
Mg$_x$Fe$_{2-x}$SiO$_4$ (see Draine 2003).
On typical sightlines
70--99\% of the Si is in grains (Jenkins 2004);
thus O$_{\rm dust}$/O$_{\rm total}\approx4\times(0.85\pm0.15)$(Si/O)$_{\rm total}\approx
4\times(0.85\pm0.15)$(Si/O)$_{\sun}\approx0.24\pm0.04$: 
the silicate grains contain
$24\pm4\%$ of the oxygen (for (Si/H)$_{\sun}=32\,$ppm 
from Asplund 2000,
and (O/H)$_{\sun}=457\,$ppm from Asplund et al.\ 2004).
The gas phase oxygen abundance tracks the total
oxygen abundance:
(O/H)$_{\rm gas}\approx (0.76\pm0.04)$(O/H)$_{\rm total}$.

\begin{figure}[!ht]
\plotfiddle{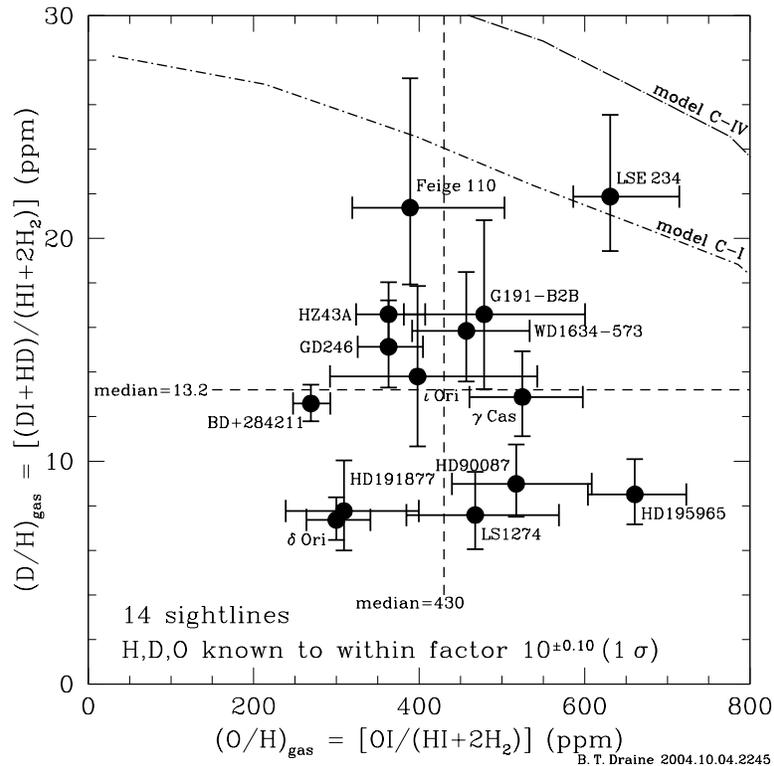}{10.0cm}{0}{50}{50}{-144}{-60}
\caption{\label{fig:DvsO}
	 (D/H)$_{\rm gas}$ vs.\ (O/H)$_{\rm gas}$ with
         1-$\sigma$ errors (see text).
	 The data 
	 do not show anticorrelation of D/H with O/H, as
	 would result from variable astration, such as
	 models C-I and C-IV from Chiappini et al.\ (2002).
	 Note that the full range of D/H is seen for
	 sightlines with (O/H)$_{\rm gas}$ between 300 and $400\,$ppm.
	 The observed variations in (D/H)$_{\rm gas}$ must be due
	 to depletion of D into interstellar dust.\newline
	 Sightlines:
	 WD~1634-573 (Wood et al.\ 2002);
	 GD~246 (Oliveira et al.\ 2002);
	 BD+284211 (Sonneborn et al.\ 2002; Spitzer et al.\ 1974);
	 Feige~110 (Friedman et al.\ 2002);
	 $\gamma$~Cas (Ferlet et al.\ 1980; Meyer et al.\ 1998; Meyer 2001);
	 $\delta$~OriA (Jenkins et al.\ 1999; Meyer et al.\ 1998; Meyer 2001);
	 $\iota$~Ori 
	 (Laurent et al.\ 1979; Meyer et al.\ 1998; Meyer 2001);
	 HD~195965 and HD~191877 (Hoopes et al.\ 2003);
	 HD~90087 (H\'ebrard et al.\ 2005);
	 HZ~43A (Kruk et al.\ 2002);
	 G~191-B2B (Lemoine et al.\ 2002); 
	 LS~1274 (Wood et al.\ 2004);
	 LSE~234 (Lecavelier des Etangs, H\'ebrard, \& Williger 2005).
	 }
\end{figure}

Figure \ref{fig:DvsO} shows tracks of D/H vs.\ O/H for two
chemical evolution models (Chiappini, Renda, \& Matteucci 2002), where the
degree of astration is a decreasing function of galactocentric radius.
These models, which include infall, have O/H varying from 300 to 600$\,$ppm as
D/H declines by only $\sim25\%$.
It is difficult to envision astration scenarios where
D/H varies by a factor of 4
without associated very
large variations 
in O/H.

However, unlike the D/H ratio, the O/H ratio does {\it not}
show large variations.
Figure \ref{fig:DvsO} shows measured values of (D/H)$_{\rm gas}$ and 
(O/H)$_{\rm gas}$ for 14 sightlines where $N({\rm HI})$, 
$N({\rm DI})$, and $N({\rm OI})$ are all reported with 1-$\sigma$
uncertainties not exceeding 0.10 dex.
For this sample, the median (O/H)$_{\rm gas}$ is $430\,$ppm
and the median (D/H)$_{\rm gas}$ is $13.2\,$ppm.
There is no indication of a
significant anticorrelation between 
(O/H)$_{\rm gas}$ and (D/H)$_{\rm gas}$:
\begin{itemize}
\item Two of the three sightlines
with the lowest values of (D/H)$_{\rm gas}$ ($\delta$~Ori and
HD~191877) have (O/H)$_{\rm gas}$ {\it below} the median.
\item The sightline with the highest (D/H)$_{\rm gas}$ in this sample
(LSE~234) has the second-highest (O/H)$_{\rm gas}$.
\item The full (factor of $\sim$3) range of variation of (D/H)$_{\rm gas}$ 
is seen among the 6
sightlines with (O/H)$_{\rm gas}$ between 300 and $400\,$ppm
($\delta$~Ori, HD~191877, $\iota$~Ori, GD~246, HZ~43A, and Feige~110).
\end{itemize}
It therefore seems very unlikely that the observed variations in 
(D/H)$_{\rm gas}$ are
due to variations in astration of the interstellar gas.
If the (D/H)$_{\rm gas}$ variations are real, some other process must
be responsible.

\subsection{Role of Dust: Depletion of D}

When we observe variations in the gas-phase abundance of, say, Si or Fe 
from one local 
sightline to another, we do not attribute it to variable astration,
but instead to variations in the fraction of the Si and Fe ``depleted''
from the gas by incorporation in
solid dust particles.
Jura (1982) suggested that even D might perhaps be depleted from the
gas phase and sequestered in
dust grains.  Since 1982 our picture of interstellar grains has
evolved to include a substantial population of very small 
polycyclic aromatic hydrocarbon (PAH) grains (see the review by Draine 2003). 
Draine (2004) argued 
that carbonaceous interstellar grains could incorporate 
enough D atoms to substantially reduce
the D abundance in the gas phase.
Under sufficiently violent conditions, the ``depleted'' D atoms
could later be released from
grains and
returned to the gas.

In this interpretation, astration has reduced the D/H abundance from
the
primordial value to (D/H)$_{\rm total}\approx23\,$ppm; incorporation of
some of the D into dust will then result in 
(D/H)$_{\rm gas}<({\rm D/H})_{\rm total}$.
The {\it variations} in the gas-phase D/H ratio are
attributed to variations from one sightline to another 
in the fraction of the D sequestered in dust grains.

\section{Hydrogen in Dust Grains}

\subsection{Observations and Grain Models\label{sec:observations}}

Measurement of the D/H ratio using ultraviolet spectroscopy is necessarily
limited to sightlines through ``diffuse'' clouds with $A_V\la2\,$mag.
Grains in clouds with $A_V\la3.3\,$mag 
do not have H$_2$O ice mantles (Whittet et al.\ 1988);
the hydrogen present in diffuse cloud 
grains must be primarily in hydrocarbons.
In fact, a broad absorption feature at 3.4$\mu$m is observed
and is identified as the C-H stretching mode in aliphatic hydrocarbons,
although the amount of material necessary to account for the observed feature
is uncertain (see, e.g., Pendleton \& Allamandola 2002).

In addition to the 3.4$\mu$m absorption feature, we also observe
emission features at 3.3, 6.2, 7.7, 8.6, and 11.3$\mu$m that are
attributed to vibrationally-excited polycyclic aromatic hydrocarbon (PAH)
molecules.  Models that quantitatively reproduce the observed emission
require substantial amounts of C in such 
particles; the model of Li \& Draine (2001) has C/H$_{\rm total} = 45\,$ppm in 
PAH grains containing $\la2500$ C atoms\footnote{%
  Here and below, H$_{\rm total}$ refers to all of the H nucleons, including
  those in the gas.}
The larger PAH grains are presumed to be clusters of smaller PAH molecules.
These might have C:H ratios in the range of 2:1 -- 4:1
(e.g., coronene C$_{24}$H$_{12}$ to
dicircumcoronene C$_{96}$H$_{24}$).  
If so, the very small PAH grains
alone may contain H/H$_{\rm total}=10$--20$\,$ppm.

How much C is in all of the grain material?
If it is assumed 
that the total interstellar C abundance is equal to current estimates
of the solar abundance 
(C/H=$246\pm23\,$ppm; Allende Prieto, Lambert, \& Asplund 2002), 
then the difference between this value and the gas-phase abundances
of CI and CII determined from absorption line measurements
allow one to estimate that dust contains
C/H$_{\rm total}=106\pm18\,$ppm (Jenkins 2004).

However, (1) solar abundances are uncertain, and 
(2) interstellar abundances may differ from solar abundances.
When one seeks to construct a grain model to reproduce the observed
extinction per H nucleon as a function of wavelength, one finds that in
addition to silicate material, it is necessary to include a
substantial amount of carbonaceous material.
The dust model of Weingartner \& Draine (2001a) assumes 
C/H$_{\rm total}\approx250\,$ppm
in dust (including the small PAH particles).
The dust models of Zubko, Dwek \& Arendt (2004) use
C/H$_{\rm total}=190$--$275\,$ppm.

Pendleton \& Allamandola (2002) estimate that about 85\% of the C in dust
is
aromatic ($sp^2$ bonding, e.g., PAHs and graphite),
and 15\% is aliphatic (``chain hydrocarbons'').
They estimate a C:H ratio $\sim$2.9:1 for the aromatic material,
and $\sim$0.46:1 for the aliphatic material.
For purposes of discussion, I will assume C/H$_{\rm total}=200\,$ppm
and (H+D)/H$_{\rm total}=70\,$ppm in aromatic hydrocarbon material, with
an additional C/H$_{\rm total}=40\,$ppm and H/H$_{\rm total}=75\,$ppm
in aliphatic material.
Because the smallest grains, which contribute most of the surface area,
appear to be PAHs, I will focus on deuteration of 
the aromatic hydrocarbons alone,
and neglect possible deuteration of the aliphatic material.

Observed values of (D/H)$_{\rm gas}$ range from $\sim$22$\,$ppm
to $\sim7\,$ppm.\footnote{%
   Some high values:
   $21.8_{-1.9}^{+2.2}\,$ppm toward $\gamma^2$~Vel (Sonneborn et
   al.\ 2000), 
   $21.4\pm4.1\,$ppm toward Feige 110 (Friedman et al.\ 2002), 
   and $21.9_{-2.8}^{+3.7}\,$ppm toward LSE 234 (Lecavelier et al.\ 2005).
   Some low values: 
   $5.0\pm1.6\,$ppm toward $\theta$~Car (Allen et al.\ 1992), 
   $7.4_{-0.9}^{+1.2}\,$ppm toward $\delta$~Ori (Jenkins et al.\ 1999),
   $7.9\pm1.9\,$ppm toward LSE 1274 (Wood et al.\ 2004).}
For dust to account for the variations in (D/H)$_{\rm gas}$,
the dust grains must be able to hold at least 
D/H$_{\rm total}\approx15\,$ppm.
If the total amount of hydrogen in the aromatic material is 
H/H$_{\rm total}\approx70\,$ppm,
then it must be possible to enrich the deuterium concentration to the
point where (D/H)$_{\rm dust} \approx 15/55 = 0.27$.
Given that the overall D/H ratio is of order $2\times10^{-5}$, processes
must act to increase the D/H ratio in the dust by 4 orders of
magnitude!

Examples of extreme deuteration are already known.
For example, fragments of the hydrocarbon material in one 
interplanetary dust particle (IDP)
analyzed in the laboratory have
D/H $\approx 0.008$ (Messenger 2000).
In some molecular 
cloud cores, extreme deuteration of small gas-phase molecules
is seen, e.g., D$_2$CO/H$_2$CO = 0.01 -- 0.1
(Ceccarelli et al.\ 2001; Bacmann et al.\ 2003),
NH$_2$D/NH$_3$ = 0.07 (Shah \& Wooten 2001), 
ND$_3$/NH$_3\approx\,$0.001 (Lis et al. 2002; van der Tak et al.\ 2002),
and CH$_2$DOH/CH$_3$OH$=0.9\pm0.3$ (Parisi et al.\ 2002).

Sandford, Bernstein, \& Dworkin (2001)
discussed 4 processes that could lead to deuterium 
enrichment in carbonaceous material:
low-temperature gas phase ion-molecule reactions;
low-temperature gas-grain reactions;
UV photon absorption followed by ejection of H;
and deuteration of carbon molecules embedded in D-rich ices.
I focus here on processes that can deuterate grains in diffuse clouds,
and show that D could be removed from the gas on timescales as short as
a few Myr.

\subsection{Energetics of Deuteration}

While H and D have the same electronic structure, and therefore the
same ``chemistry'', the C-D bond is stronger than the C-H bond because
of the reduced zero-point energy.  It is easy to estimate the magnitude of
this energy difference.  
To be specific, let us consider PAH molecules, since we have direct
evidence that at least some of the solid carbon is in this form.

PAHs show three emission features associated with vibrations of the C-H bond:
the C-H stretch at $\lambda_1\approx3.3\mu$m, in-plane C-H bending at
$\lambda_2\approx8.6\mu$m, and out-of-plane C-H bending at 
$\lambda_3\approx11.3\mu$m.
The stretching and bending modes can be approximated as harmonic oscillators.
If the H is replaced by D, the reduced mass of the C-H oscillator
will increase by a factor 13/7, and the frequency will
be reduced by a factor $\sim\sqrt{7/13}$.
Each of the vibrational modes has zero-point energy $\hbar\omega/2$
associated with it, so (summing over the three modes) 
the zero-point energy will be lowered by 
\begin{equation}
\Delta E({\rm CH} - {\rm CD}) \approx 
\frac{1}{2}\left(1-\sqrt{\frac{7}{13}}\right)
\sum_{j=1}^3 \frac{hc}{\lambda_j}
= 0.083{\rm eV} = k\times 970{\rm K}
~~~.
\end{equation}
Therefore it is energetically favorable for the H in PAHs to be replaced
by D.  If atomic H and D in the gas were in thermodynamic equilibrium with
deuterated PAH material at temperature $T_d$, we would expect to have
\begin{equation} \label{eq:condition on Tdust}
\left(\frac{\rm D}{\rm H}\right)_{\rm PAH} <
\left(\frac{\rm D}{\rm H}\right)_{\rm gas}e^{\Delta E/kT_d}
~~~,
\end{equation}
where the inequality is because we may not reach a steady-state,
and there may be other processes acting to return D to the gas.

Now suppose that (D/H)$_{\rm total}$=22$\,$ppm.  If the aromatic grains
contain
D/H$_{\rm total}=15\,$ppm and H/H$_{\rm total}=55\,$ppm,
eq.\ (\ref{eq:condition on Tdust}) becomes
\begin{equation}
\frac{15}{55} < 7\times10^{-6}e^{970{\rm K}/T_d}
~~~,
\end{equation}
or $T_d < 970{\rm K}/\ln\left[15/(55\times7\times10^{-6})\right] = 91$K.
The ISM is, of course, far from LTE, so thermodyamic arguments are
suspect.  
Nevertheless, since interstellar grain material is generally at 
temperatures $T_d\la20{\rm K}\ll91$K,
we can at least contemplate the possibility that PAH grains
exposed to atomic H and D might attain D:H ratios as high as $\sim$0.3.

It can easily be shown that thermal desorption also cannot be important.
At such temperature $T_d < 91\,$K, the rate for
thermal desorption of chemically-bound H in PAHs
is entirely negligible.
Therefore it does not appear that the required degree of deuteration
can be achieved by the thermal desorption process discussed by
Allamandola, Tielens, \& Barker (1989).

Because grains can catalyze H$_2$ formation,
it is important to note that
the energy difference
\begin{equation}
\Delta E ({\rm H}_2 - {\rm HD}) \approx
\frac{1}{2}
\frac{hc}{2.4\mu{\rm m}}
\left(1-\sqrt{\frac{3}{4}}\right) = 0.035{\rm eV}
~~~,
\end{equation}
is 0.048~eV smaller than $\Delta E({\rm CH} - {\rm CD})$, so that
it is energetically favorable to leave the D attached to C and instead
abstract an H to form H$_2$ rather than HD
(2.4$\mu$m is the wavelength of H$_2$1-0 Q branch lines).
Thus even if the H$_2$ formation process involves hydrogen
abstraction,
H$_2$ will be strongly favored over formation of HD, leaving D atoms
chemically bound
in the hydrocarbon grain.
This differs from the prediction by Lipshtat, Biham,
\& Herbst (2004) of enhanced HD and D$_2$ formation if H and D are only
physisorbed on the grain surface.

Now the key question is: can we imagine a kinetic pathway whereby
extreme deuteration of grains might come about?

\subsection{Kinetics of Deuteration}

While it is known that PAHs can form in the high-temperature chemistry of
flames, the processes responsible for the formation of 
interstellar PAHs and other carbonaceous material remain unclear.
We will assume that at the time of their formation, the carbon solids 
are minimally deuterated, incorporating H and D approximately in proportion to 
gas-phase abundances.  Given that carbonaceous grains 
incorporate only $\sim$145$\,$ppm
of all of the hydrogen, a negligible fraction of the D would initially
be trapped in PAHs.

For grains to become heavily deuterated, 
it is necessary that 
\begin{enumerate}
\item D or D$^+$ arriving at a carbonaceous grain must have a
significant probability of being incorporated into the grain;
\item D that has been incorporated must not be removed by exchange
reactions with arriving H or H$^+$ -- if this were to happen with
any appreciable probability, the steady-state abundance of D in the
grain would be very low.
\end{enumerate}

\begin{figure}[!ht]
\plotfiddle{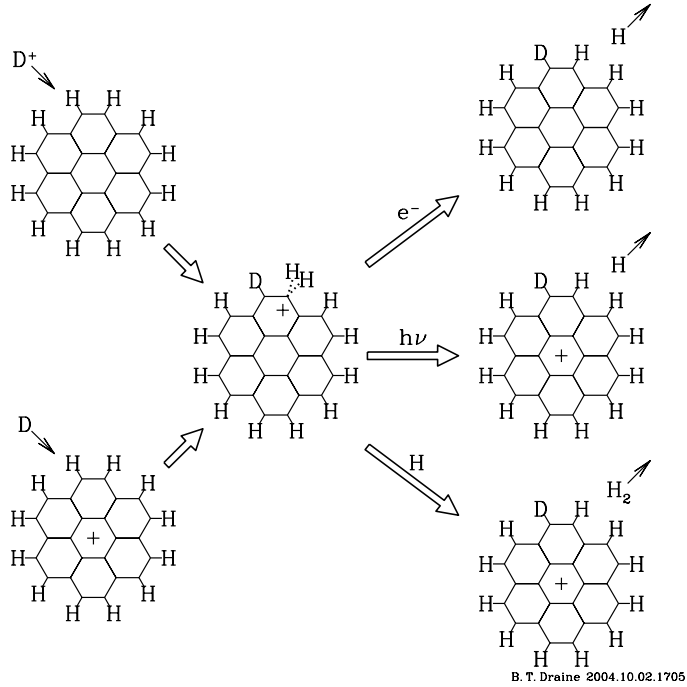}{9.0cm}{0}{45}{45}{-144}{-60}
\caption{\label{fig:react}
         Examples of reaction pathways for deuteration of a generic PAH 
	 (see text), by collision of D$^+$ with a neutral PAH (upper left)
	 or collision of D with a PAH$^+$ cation (lower left) to create
	 a hydro-PAH cation, with two H atoms bonded to one C.  
	 Subsequent reaction with incoming H would
	 remove one of the extra H atoms and 
	 form H$_2$.  Other pathways also exist. 
	 }
\end{figure}

We consider conditions in cold HI clouds (the ``cold neutral medium'', or CNM),
with density $n_{\rm H} \approx 30 \,{\rm cm}^{-3}$,
electron density $n_e\approx 0.03\, {\rm cm}^{-3}$, with perhaps
1/3 of the electrons contributed by ionized metals, and 2/3 by
ionized hydrogen.  Thus the hydrogen ionization fraction
$x_{\rm H}\approx 7\times10^{-4}$; because of rapid charge exchange,
deuterium will have the same ionization fraction.
We take the local average starlight background.

Impinging low-energy D atoms are probably unable to
react chemically with an already saturated neutral hydrocarbon -- the
arriving D will most likely be physisorbed or chemisorbed on the
surface of the neutral PAH, from which it can subsequently be removed by
reacting with an arriving H to form HD, or by thermal desorption following
heating of the PAH by photon absorption.

As discussed by Bauschlicher (1998) for the
naphthalene cation C$_{10}$H$_8^+$, a D atom can
react with a PAH cation, producing a hydro-PAH cation,
with two hydrogens sharing a single carbon (see Figure \ref{fig:react}).
How rapidly will this occur?

It is not known if
thermal D atoms will react with PAHs much larger than
naphthalene if they are only singly
or doubly charged, so let us consider only PAH cations containing
$<100$ C atoms.
Collision cross sections are estimated following Draine \& Sutin (1987), 
the PAH size distribution is from Li \& Draine (2001) 
and Weingartner \& Draine (2001a),
and
the PAH charge distribution is estimated 
following Weingartner \& Draine (2001b).
For a neutral D in the CNM, I find the rate to collide and react with PAH
cations containing $< 100$ atoms is
$1.3\times10^{-14}{\rm s}^{-1}$; if the probability of reaction is of
order unity, and there is no process returning D to the gas phase,
then the gas-phase
D abundance in the CNM would decline by a factor $1/e$ in only $\sim2.4\,$Myr.

Another pathway begins with
D$^+$ colliding with a neutral (or negative) 
PAH; 
the D$^+$ will arrive at the PAH with sufficient kinetic energy
that atomic rearrangements can take place among the peripheral hydrogens
near the point of impact.
The excess energy will be
thermalized in the vibrational modes; the collision complex is expected
to maximize its binding energy by an exchange reaction that would
replace a C-H bond with a C-D bond, thus incorporating the D
and creating a hydro-PAH cation just as for the case of a neutral D
colliding with a PAH cation.  
It seems likely that every arriving D$^+$ ion will be incorporated
into the PAH via this pathway, thus deuterating the PAH and removing
D from the gas.

The rate for species $X$ to collide with a grain can be written 
$n_{\rm H}\alpha_g(X)$,
where $\alpha_g(X)$ depends on the PAH charge distribution.
Ions have relatively high rates for collisions with neutral PAHs because
of the induced polarization interaction, and even higher rates for
negative PAH ions because of Coulomb focussing.
Weingartner \& Draine (2001c) estimated
$\alpha_g({\rm H}^+)\approx 6\times10^{-14}{\rm cm^3\,s}^{-1}$ in the CNM;
$\alpha_g({\rm D}^+)$ will be smaller
by $1/\sqrt{2}$,
$\alpha_g({\rm D}^+)\approx4\times10^{-14}{\rm cm^3\,s}^{-1}$.  
Thus the rate for D nuclei to react with a grain by this pathway is
$x_{\rm H} n_{\rm H} \alpha_g({\rm D}^+)\approx 8\times10^{-16}{\rm s}^{-1}$.
This is a factor of 16 slower than the rate for D nuclei to react with
PAH cations, and therefore is of only secondary importance.
Together, the gas-phase processes appear able to remove D from the
gas, and deuterate PAHs, on a time scale of $\sim2.3\,$Myr.

Evolution of the grain population is not well understood, but it appears that
the larger grains move relative to 
the gas with speeds of $\sim1\,{\rm km\,s}^{-1}$
resulting from MHD turbulence (Yan et al.\ 2004).  This can sweep up the
PAHs on $\sim10^7$yr timescales 
and incorporate them into larger grains; occasional shattering
of larger carbonaceous grains when they collide replenishes the PAH population.
In this scenario, all of the aromatic grain material eventually becomes
enriched with D, not just the fraction that is instantaneously
present in small free-flying PAHs.

The proposed extreme deuteration of PAH material imposes strong
requirements on branching ratios for reactions of PAH material with
impinging H and D atoms and ions, and is therefore subject to
laboratory test.
In particular, an experiment with low energy D$^+$ impinging on
partially-deuterated PAH material should have a high probability of
incorporating the D into the solid, while reactions of low energy H$^+$ with
the same sample should have a very small probability of abstracting D
from the sample.

\section{Return of D to the Gas Phase}

Grains don't last forever -- sputtering and grain-grain
collisions following the passage of $v_s > 100\,{\rm km\,s}^{-1}$
shock waves can return atoms -- including D -- from the solid to the gas phase.
When this occurs, (D/H)$_{\rm gas}$ should increase, along with the
gas-phase abundances of other elements, such as Si, Ti, and Fe, that are
normally strongly depleted.  The hypothesis that dust is responsible
for the observed variations in (D/H)$_{\rm gas}$ therefore predicts
that the gas-phase abundances of
Si, Ti, Fe and other refractory metals should show a positive correlation
with the gas-phase abundance of D.  

This prediction appears to 
be confirmed by a recent study finding
a positive correlation 
between (D/H)$_{\rm gas}$
and (Fe/H)$_{\rm gas}$
(Linsky et al.\ 2005).

\section{Can D in Grains be Observed?}

\subsection{Spectroscopy}

The emission features at 3.3, 8.6, and 11.3$\mu$m provide direct spectroscopic
evidence for the presence of C-H bonds in vibrationally-excited interstellar
PAH material.
Can C-D bonds be observed through corresponding emission features?
The C-D in-plane bending mode
at $\sim1.36\times8.6\mu{\rm m}=11.7\mu{\rm m}$ would be confused with the
C-H out-of-plane bending modes at $\sim11.3\mu{\rm m}$,
and the out-of-plane bending mode at 
$\sim1.36\times11.3\mu{\rm m}=15.4\mu{\rm m}$ falls in a region where
other PAH features (C-C skeleton modes) are present.
The best hope appears to be the C-D stretching mode
at $\sim1.36\times3.3\mu{\rm m}\approx4.5\mu{\rm m}$.
In fact, Peeters et al.\ (2004) report detection of weak emission features
at 4.4$\mu$m and 4.68$\mu$m from the Orion Bar and M17 photodissociation
regions.  
Identification of the features as due to C-D is uncertain, but
if the features are indeed due to C-D modes, Peeters et al.\
estimate (D/H)$_{\rm PAH}=0.17\pm0.03$ in Orion and $0.36\pm0.08$ in M17.
These D/H values are just in the range that we have estimated above to
be necessary to lower (D/H)$_{\rm gas}$ to the observed values!

Additional spectroscopy of the emission from 
PDRs in the 4--5$\mu$m region to confirm these detections and identifications
would be of great
value.  Unfortunately, the IRS instrument on Spitzer only works longward
of 5.2$\mu$m.

Detection of absorption features produced by deuterated PAH material is
unpromising, because 
absorption by the C-H stretching mode in aromatic hydrocarbons
is quite weak -- even the 3.3$\mu$m aromatic C-H stretch 
feature has not been seen in absorption.
Detection of the 4.5$\micron$ C-D stretch in absorption does not appear
to be possible.
The C-D bending modes will also be too weak to detect, and in any case
would be confused with other PAH modes, as discussed above.

\subsection{Interstellar Grains in the Laboratory}

The local interstellar medium appears to have
(D/H)$_{\rm gas}\approx 15\,$ppm (see Fig.\ 6 of Wood et al.\ 2004).  
If the total D/H $\approx23\,$ppm, then the local dust grains would be
deuterated.  
For our above provisional value of (H+D)$_{\rm dust} \approx 145\,$ppm
(including aliphatic hydrocarbons), 
we would estimate (D/H)$_{\rm dust}=8/145\approx0.06$.
The {\it Stardust} mission (Brownlee et al.\ 1994)
should have captured interstellar dust grains
in its aerogel dust collectors, to be returned to earth in January 2006.
If it has collected grain material that is representative
of the interstellar grain population, this would be expected
to have D/H $\approx 0.06$ if dust is responsible for the observed variations
in (D/H)$_{\rm gas}$; the aromatic component of such grains, if isolated, 
might have
an even higher D/H ratio.  Laboratory measurements of D/H in
the returned interstellar grain samples will test this hypothesis.

As noted above (\S\ref{sec:observations}), IDPs
are sometimes seen to be strongly deuterated.
The largest D/H ratio seen
thus far is $0.008$ (Messenger 2000).  If interstellar grains have 
D/H ratios as large as $0.3$, as proposed here, why are these not seen
in IDPs that are thought to originate in comets?

First, while the median (D/H)$_{\rm gas} \approx 14\,$ppm
in Figure \ref{fig:DvsO}, two of the 14 sightlines
have (D/H)$_{\rm gas} \ga 20\,$ppm.
Presumably this material
was shocked relatively recently.
Perhaps the solar system formed
out of such matter 4.5$\,$Gyr ago.
Second, while comets are thought to be made of relatively primitive
material, it is possible that accretion into the solar nebula might
involve a phase with sufficiently high temperatures and densities for
significant exchange of
H and D with the gas to take place, thereby lowering the D/H ratio
in the grains.  

The observed D/H ratios in interplanetary dust particles are compatible
with the hypothesis that interstellar grains can reach (D/H) values
as large as $\sim0.3$ .

\section{Summary}

\begin{enumerate}
\item The lack of correlation between (D/H)$_{\rm gas}$ and (O/H)$_{\rm gas}$
in Figure \ref{fig:DvsO} argues strongly against variations in astration
as the explanation for the observed variation in (D/H)$_{\rm gas}$.
\item It appears possible for dust grains to remove D from the gas phase
in amounts sufficient to account for the observed variations in
(D/H)$_{\rm gas}$ from one sightline to another.
\item Collisions of D with PAH cations, and collisions of D$^+$ with PAHs, 
are expected to result in incorporation of
D into the PAH.  The rate of such collisions is sufficient to deplete
D from the gas on time scales of $\sim$2~Myr in cool H~I clouds.
\item Grain destruction should return D to the gas along with other
elements found in refractory grains.
There should therefore be an observable 
correlation between (D/H)$_{\rm gas}$ and
the gas-phase abundances of Mg, Si, Ca, Ti, Fe, and other elements that
exhibit strong depletions.
This prediction appears to be confirmed by a recent study finding
a positive correlation between (D/H)$_{\rm gas}$
and (Fe/H)$_{\rm gas}$ (Linsky et al.\ 2005). 
\item Carbonaceous interstellar material collected by {\it Stardust},
if representative of grains from the local interstellar medium,
should be strongly deuterated.
\item Laboratory studies can test whether the proposed deuteration
reactions have the required branching ratios.
\end{enumerate}

\acknowledgements 
I am grateful to 
the Scientific Organizing Committee for invitating me to participate.
I thank J. Black, G. H\'ebrard and J.L. Linsky for helpful discussions,
and R.H. Lupton for making available the SM plotting package.
This work was supported in part by NSF grant AST-9988126 and by grants from the
W.M. Keck Foundation and the Monell Foundation.




\end{document}